\documentclass[aps,pra,amsfonts,nofootinbib,preprint,letterpaper,showpacs,linenumbers]{revtex4-1}

\usepackage{xspace}
\usepackage{graphicx}
\usepackage{dcolumn}
\usepackage{braket}
\usepackage{srcltx}

\newcommand{\figab}{Fig.~}
\newcommand{\tabab}{Table~}
\newcommand{\secab}{Sec.~}
\newcommand{\eqab}{Eq.~}
\newcommand{\citab}{Ref.~}
\newcommand{\citsab}{Refs.~}

\newcommand{\eqref}[1]{(\ref{#1})}

\newcommand{\abini}{\textit{ab~initio}\xspace}

\newcommand{\stgnd}{1s$^2$\xspace}
\newcommand{\sttstp}{2s2p ($^1\mathrm{P^o}$)\xspace}
\newcommand{\sttst}{2s$^2$ $(^1\mathrm{S^e})$\xspace}
\newcommand{\sttpt}{2p$^2$ $(^1\mathrm{S^e})$\xspace}

\begin{document}
  \title{Femtosecond transparency in the extreme ultraviolet}
  \author{\firstname{Michal} \surname{Tarana}}
  \email{michal.tarana@jila.colorado.edu}
  \author{\firstname{Chris} H. \surname{Greene}}
  \affiliation{Department of Physics and JILA, University of Colorado, Boulder, Colorado 80309-0440, USA}
  \begin{abstract}
    Electromagnetically induced transparency-like behavior in the extreme ultraviolet (XUV) is studied theoretically, including the effect of intense 800\thinspace nm laser dressing of He \sttstp ($^1\mathrm{P^o}$) and \sttpt ($^1\mathrm{S^e}$) autoionizing states. We present an \abini solution of the time-dependent Schr\"odinger equation (TDSE) in an \textsl{LS}-coupling configuration interaction basis set. The method enables a rigorous treatment of optical field ionization of these coupled autoionizing states into the $N = 2$ continuum in addition to $N = 1$.  Our calculated transient absorption spectra show encouraging agreement with experiment.
  \end{abstract}
  \maketitle
  \section{Introduction}
  The use of two laser fields to dress the eigenstates of a three-level system has attracted significant attention over past two decades~\cite{Cohen2011}. The interaction of light with a coherent atomic ensemble leads to important quantum interference phenomena, including electromagnetically induced transparency (EIT)~\cite{Harris1990,marangos-rmp}, in which the laser-induced coherence of atomic states leads to quantum interference between the excitation pathways that control the optical response. It can be used to manipulate the linear and nonlinear susceptibilities of matter from the near-infrared to ultraviolet. This has resulted in applications such as slowed and stored light as well as nonlinear frequency conversion~\cite{Lukin2003}.
  
  Presently there is great interest in the application of light in the X-ray regime, produced by high-order harmonics and free electron lasers, to investigate novel coherent X-ray optical phenomena. In this context, ultrafast electron dynamics of the autoionization processes in atoms has gained attention in recent years. The advent of femtosecond and attosecond XUV laser pulses in a pump-probe setting has made it possible to follow these processes in real time~\cite{Wickenhauser}. These experiments involving two coherent laser pulses allow study of the time-resolved interaction of the direct ionization pathway with pathways involving one or more resonant states. This opens the possibility of new ways to monitor and control the autoionization dynamics~\cite{Wang-Chini-Ar,Argenti2010} and consequently the XUV absorption properties of the dressed medium. 
  
  Theoretical studies due to \citet{Bachau86} and \citet{Themelis2004} predicted the transformation of the original unperturbed Fano line-shape~\cite{fano-ci} into an Autler-Townes doublet due to Rabi flopping between a pair of coherently-coupled doubly-excited states. \citet{Loh2008} observed EIT-like behavior in an XUV probe induced by coherent coupling of the \sttstp and \sttpt doubly-excited states in helium, where the XUV probe was created from IR laser-produced high-order harmonics. Specifically, that study used a femtosecond infrared (IR) pulse both to couple the doubly-excited states and to produce the probe pulse. \citet{Buth-Santra-EIT} predicted an EIT for X-rays in laser dressed neon. More recently, \citet{Ranitovic2011} used a two-color multiphoton ionization of helium in combined IR laser and XUV high harmonic fields to control the transparency. They showed that this scheme can induce full electromagnetic transparency.
  
  Quantum mechanical calculations of the photoabsorption spectra for the dressed atoms represent a challenge for theory, especially when the dressing laser pulse couples doubly-excited states. The time-consuming solution of the time-dependent Schr\"odinger equation (TDSE) is required for every photon energy of the probe pulse. An alternative theoretical treatment has been introduced by~\citet{Tong2010} and successfully used in \citab\cite{Ranitovic2011}. It is based on the Floquet representation and on the calculation of the autocorrelation function that is indepdendent of the probe laser photon energy. Therefore, the time and memory demanding numerical calculation of the  autocorrelation function is performed only once. Since the photoabsorption cross section has a simple analytical dependence on the autocorrelation function, its calculation is not difficult. However, the approach introduced in \citab\cite{Tong2010} involves a diagonalization of the Hamiltonian matrix in the Floquet representation, typically having dimensions $10^6\times 10^6$ and this step is still challenging for a supercomputer.
  
  To our knowledge, there is no method which enables a routine theoretical calculation of the photoabsorption spectra in the presence of a strong dressing laser pulse. The aim of the present work is to make a step towards the improvement of the theoretical treatment of this problem. This study deals with the modified Fano line-shape in the photoabsorption spectrum of the He \sttstp doubly-excited state caused by the coupling of this state with the \sttpt doubly-excited state by an IR femtosecond pulse. \citet{Loh2008} studied this effect experimentally and interpreted obtained results as EIT-like behavior. A simplified three-level\cite{Madsden2000} model treatment of this problem was included in \citab\cite{Loh2008}. The model describes formation of an Autler-Townes doublet and takes into account the effect of the optical field ionization of the doubly-excited states due to the IR dressing laser using a somehow \textit{ad hoc} adaptation of the Ammosov-Delone-Krainov (ADK) method. Our approach presented in this work is based on the numerical solution of the TDSE. It takes into account the possible coupling of more than two autoionizing states and the continuum in principle more accurately than the approach presented in \citab\cite{Loh2008}. Therefore, the present treatment of the optical field ionization of the doubly-exited states due to the coupling laser field is more quantitative as well. Photoabsorption spectra calculated using parameters corresponding to the experimental setup used in \citab\cite{Loh2008} are compared with results obtained for reduced IR pulse intensity and for significantly reduced IR detuning. Comparison of all these results allows us to study the effect of the optical field ionization and of the IR laser detuning on the formation of EIT-like structures in the spectra.
  
  The rest of this paper is organized as follows: The theoretical approach and assumptions of our model are discussed in \secab\ref{sec:theo}. A detailed description of the field free Hamiltonian and static photoabsorption spectrum is given in \secab\ref{sec:staticfree}, while \secab\ref{sec:dressed} discusses the photoabsorption yields of the dressed atom.
  
  Atomic units are used throughout the rest of the paper.
  \section{Theoretical Approach}
  \label{sec:theo}
  The model introduced in the present study is based on the experimental publication of to \citet{Loh2008}. In that experiment the \sttstp($^1$P$^\mathrm{o}$) and \sttpt($^1$S$^\mathrm{e}$) doubly-excited states of He are coupled by a strong 800\thinspace nm field. The \sttstp and \sttpt states are located at 60.15~\cite{Domke1996} and 62.06\thinspace eV~\cite{Burgers1995} above the \stgnd ground state, respectively. This system is then probed after a variable time delay by an XUV pulse. The XUV probe pulse is produced with a table-top, laser-based setup by high-order harmonic generation (HHG). The authors use a commercial Ti:sapphire laser system (2.4\thinspace W, 800\thinspace nm, 42\thinspace fs, 1\thinspace kHz) to produce the optical dressing and HHG beam. The estimated photon flux at the source is $10^5$ photons per pulse for the high-order harmonic centered at 60.2\thinspace eV. The XUV pulse duration is estimated to be 30\thinspace fs FWHM. The dressing pulse peak intensity is $1.4\times10^{13}$\thinspace W/cm$^2$ and its duration is 42\thinspace fs FWHM. The spectrometer used in~\cite{Loh2008} measures the transmission spectrum instead of the absorption spectrum. The unknown fraction of dressed atoms in the interaction volume prevents the photoabsorption spectrum of the dressed atom from being retrieved experimentally. Therefore, authors measure the transient absorption spectra and identify EIT-like structures.
  
  The XUV laser pulse couples the \stgnd ground state with the \sttstp autoionizing state with energy 35.56\thinspace eV above the $N=1$ ionization threshold. Therefore, it is reasonable to assume that every absorption of an XUV photon leads to photoionization. The aim of the present work is to calculate the population of the ground state at the end of the pulse as a function of the XUV photon energy $\hbar\omega_X$ for different dressing-probe time delays $t_D$. The computed loss of the ground state population is interpreted as the total photoabsorption yield. Its comparison with the yields obtained when the IR pulse is absent allows for recognition of EIT-like structures.

  Helium is a two-electron system. The two-particle basis set used in the present calculations consists of $LS$-coupled independent-particle basis functions~\cite{Aymar1996, LagmagoKamta2002}. The Hamiltonian for the interaction of each electron with the core is
  \begin{equation}
    h(\mathbf{r})=-\frac{1}{2}\nabla^2-\frac{2}{r},
  \end{equation}
  where $\mathbf{r}$ is the coordinate of the electron with respect to the nucleus. The corresponding (radially rescaled) single-particle radial wave function $\varphi(r)$ for an electron with orbital angular momentum $l$, is given by
  \begin{equation}
    \left(-\frac{1}{2}\frac{d^2}{dr^2}+\frac{l(l+1)}{2r^2}-\frac{2}{r}\right)\varphi(r)=\varepsilon\varphi(r).
    \label{eq:oneorbs}
  \end{equation}
  The atomic Hamiltonian including the interaction between the electrons is
  \begin{equation}
    H=h(\mathbf{r_1})+h(\mathbf{r_2})+\frac{1}{r_{12}},
  \end{equation}
  where $r_{12}=\left|\mathbf{r}_1-\mathbf{r}_2\right|$ is the distance between the electrons.
  
  To calculate the time development of the ground state population we solve the TDSE for the two linearly-polarized laser fields in the length gauge:
  \begin{equation}
    i\frac{\partial}{\partial t}\Psi(\mathbf{r}_1,\mathbf{r}_2,t)=\left[H+D_I(t + t_D)+D_X(t)+V_C\right]\Psi(\mathbf{r}_1,\mathbf{r}_2,t),
    \label{eq:alltdse}
  \end{equation}
  where $D_{I,X}(t)=\mathbf{E}_{I,X}(t)\cdot\mathbf{d}=\mathbf{E}_{I,X}(t)\cdot(\mathbf{r}_1+\mathbf{r}_2)$ are the interaction operators for the IR and XUV laser pulse in the dipole approximation, respectively. $V_C$ denotes the complex absorbing potential (CAP) added to treat ionization properly within the context of a finite volume calculation.
  As in the experimental convention, a positive time delay $t_D$ means that the XUV probe pulse center arrives later than the pump pulse center.
  The electric fields $\mathbf{E}_I(t)$ and $\mathbf{E}_X(t)$ are given by
  \begin{equation}
      \mathbf{E}_{I,X}(t)=\mathbf{\hat{z}}E_{I,X}(t)=\mathbf{\hat{z}}E_{I,X}^0f(t)\cos\left(\omega_{I,X}t\right),
  \end{equation}
  where $\omega_{I,X}$ are the IR and XUV laser field frequencies, $\mathbf{\hat{z}}$ is the unit vector along the polarization axis, $E_{I,X}^0$ are the peak amplitudes of the electric fields and $f_{I,X}(t)$ are the pulse envelopes. Throughout this work it is assumed that both envelopes have the $\cos^2$ form
  \begin{equation}
    f_{I}(t)=\left\{
    \begin{array}{ll}
      \cos^2\left(\pi t/\tau_{I}\right), & -\tau_{I}/2\leq t\leq\tau_{I}/2\\
      \sqrt{0.13}\cos^2\left(\pi (t-t_P)/\tau_{IP}\right), & -\tau_{IP}/2\leq t-t_P\leq\tau_{IP}/2\\
      0, & \text{otherwise,}
    \end{array}
    \right.
    \label{eq:irenvelop}
  \end{equation}
  resp.,
  \begin{equation}
    f_{X}(t)=\left\{
    \begin{array}{ll}
      \cos^2\left(\pi t/\tau_{X}\right), & -\tau_{X}/2\leq t\leq\tau_{X}/2\\
      0, & \text{otherwise,}
    \end{array}
    \right.
  \end{equation}
  where $\tau_{I,X}$ denote the durations of the respective laser pulses. The second term in \eqab\eqref{eq:irenvelop} denotes the IR postpulse which appears in the experiment~\cite{Loh2008} with time delay $t_P=88$\thinspace fs and duration $\tau_{IP}=30$\thinspace fs.
  The CAP used in the present calculations has the form~\cite{Riss1993}
  \begin{equation}
    V_C=\left\{
    \begin{array}{ll}
      0, & r_1\leq r_b\text{ and }r_2\leq r_b\\
      -i\eta\left[(r_1-r_b)^2+(r_2-r_b)^2\right], & \text{otherwise,}
    \end{array}
    \right.
    \label{eq:capop}
  \end{equation}
  where $\eta>0$ is the strength of the CAP and $r_b$ is the inner boundary of the absorber. The TDSE~\eqref{eq:alltdse} is integrated in a box using the expansion of the time-dependent wave function $\Psi(\mathbf{r}_1,\mathbf{r}_2,t)$ in a basis set consisting of eigenstates of the field-free atomic Hamiltonian $H$:
  \begin{equation}
    \Psi(\mathbf{r}_1,\mathbf{r}_2,t)=\sum_{L=0}^{L_{\mathrm{max}}}\sum_{n=1}^{N_L}c_{Ln}(t)\Phi_{Ln}(\mathbf{r}_1,\mathbf{r}_2),
    \label{eq:timexpand}
  \end{equation}
  where $\Phi_{Ln}(\mathbf{r}_1,\mathbf{r}_2)$ denotes the n\textsuperscript{th} eigenfunction of $H$ having total angular momentum $L$ and corresponding to the real ``box'' eigenenergy $\epsilon_{Ln}$:
  \begin{equation}
    H\Phi_{Ln}(\mathbf{r}_1,\mathbf{r}_2)=\epsilon_{Ln}\Phi_{Ln}(\mathbf{r}_1,\mathbf{r}_2).
    \label{eq:eigstates}
  \end{equation}
  The expansion~\eqref{eq:timexpand} includes $N_L$ eigenstates for the angular momentum $L$, and it includes all the angular momenta up to $L_{\text{max}}$.
  The basis set used in the variational calculations of the field-free eigenstates $\Phi_{Ln}(\mathbf{r}_1,\mathbf{r}_2)$ consists of \textsl{LS}-coupled independent-particle basis functions. Antisymmetric two-electron basis functions coupled to form a state of definite $S$, $L$, and parity $P$ are constructed in terms of one-electron orbitals~\eqref{eq:oneorbs} as explained in~\citsab\cite{Aymar1996, LagmagoKamta2002}. The radial one-electron wave functions $\varphi_{jl}(r)$ are obtained by solving \eqab\eqref{eq:oneorbs} in a radial box of size $r_0>r_b$ with boundary conditions $\varphi_{jl}(0)=\varphi_{jl}(r_0)=0$ for both positive and negative energies $\varepsilon$. The index $j$ denotes the number of nodes of the radial wave function inside the box. \eqab\eqref{eq:oneorbs} is solved by expansion of the radial orbitals in the basis set of \textsl{B}-splines~\cite{Bachau2001} $B_i^k(r)$ satisfying the bound-state boundary conditions:
  \begin{equation}
    \varphi_{jl}(r)=\sum_{i,k}c_{jlik}B_i^k(r).
  \end{equation}
  The radial box is divided into several intervals (denoted by index $k$) by knots and the \textsl{B}-splines are defined in terms of the knots in each interval. 
  
  Use of the \textsl{LS}-coupled basis as discussed above is nowadays a standard technique frequently employed in the theoretical treatment of both time-dependent~\cite{LagmagoKamta2002, McCurdy2004, Bachau2001} and time-independent~\cite{Aymar1996} problems for two active electron atomic systems. The radial one-electron orbitals are often expressed in terms of \textsl{B}-splines~\cite{LagmagoKamta2002, McCurdy2004, Bachau2001} or discrete variable representation (DVR) basis sets~\cite{Palacios2008}.
  
  In order to suppress the reflections from the outer boundary of the radial box in the solution of the TDSE~\eqref{eq:alltdse} and to keep its size reasonably small, the CAP term~\eqref{eq:capop} is added to the Hamiltonian used for the time-propagation. The presence of this term makes the total Hamiltonian of the system a complex symmetric rather than hermitian operator~\cite{Riss1993}. One note should be made at this point regarding the form of the matrix elements of all the operators used here. Since all the calculations are performed in the eigenrepresentation of the field-free atomic Hamiltonian $H$ not including the CAP, the operators $H$, $D_I(t)$ and $D_X(t)$ have real matrix elements, as it is convenient in hermitian quantum mechanics. $V_C$ is the only non-hermitian term in the total Hamiltonian. Therefore, it is not necessary to distinguish the standard hermitian scalar product and complex symmetric product throughout the present work (see~\citab\cite{moiseyev-rep} and references therein for details).
  
  Both laser pulses are linearly polarized along the $z$-axis, and of course the $^1\mathrm{S^e}$ ground state of He has zero projection $M_L$ of the total orbital angular momentum on the polarization axis. Accordingly it is also enough to consider only the $z$-component of the dipole operator. According to the selection rules for the matrix elements of the dipole moment operators $\left(\mathbf{d}_z\right)_{LnL'n'}=\bra{\Phi_{Ln}}z_1+z_2\ket{\Phi_{L'n'}}$, these are non-zero only if $L-L'=\pm1$~\cite{Sobelman1972}.
  
\begin{figure}
    \includegraphics[scale=0.65]{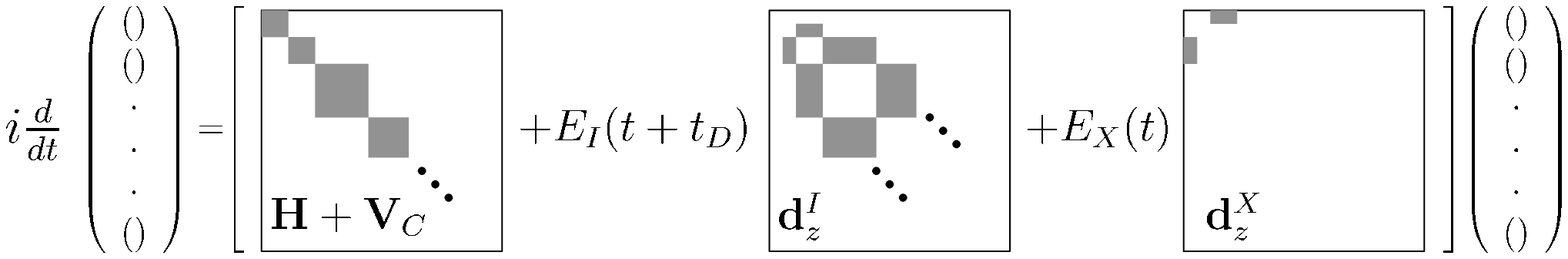}
        \caption{Structure of the matrices in the system of coupled equations~\eqref{eq:approxmat}}. The grey areas denote the non-zero blocks corresponding to definite total angular momenta and their dipole coupling. Matrix $\mathbf{d}_z^I$ has zero first line and first column, while matrix $\mathbf{d}_z^X$ has non-zero only the first row and column.
    \label{fig:eqstr}
  \end{figure}

  Projection of the TDSE~\eqref{eq:alltdse} on the eigenstates of the field-free Hamiltonian~\eqref{eq:eigstates} yields the following set of coupled differential equations of the first order:
  \begin{equation}
      i\mathbf{\dot{c}}(t)=\left[\mathbf{H}+\mathbf{V}_C+\mathbf{D}_I(t+t_D)+\mathbf{D}_X(t)\right]\mathbf{c}(t),
      \label{eq:tdmat}
  \end{equation}
  where $\mathbf{c}(t)$ is the vector of expansion coefficients used in~\eqref{eq:timexpand} and the matrices on the right hand side are the representations of the corresponding operators in the basis set $\Phi_{Ln}(\mathbf{r}_1,\mathbf{r}_2)$ introduced above. The dimension of this system of differential equations is $N=\sum_{L=0}^{L_{\text{max}}}N_L$. In the present representation $\mathbf{H}$ is a real diagonal matrix of the field-free atomic eigenenergies. Since $V_C$ given by \eqab\eqref{eq:capop} has no angular dependence, $\mathbf{V}_C$ is an imaginary and symmetric block diagonal matrix with $L_{\text{max}}+1$ non-zero blocks having dimensions $N_L$. $\mathbf{D}_{I,X}(t)$ are real symmetric banded matrices with only non-zero blocks corresponding to the coupling of the total angular momentum $L$ with $L\pm1$, because we only consider $M_L=0$ states in this study.
  The physical picture of the phenomenon studied here suggests that the IR dressing laser pulse couples the singlet autoionizing states \sttstp, \sttpt and the continuum states made accessible by the IR pulse. The energy difference between the ground state and the lowest odd-parity, singlet excited state - 1s2p($^1$P$^o$) of the field-free atom is so high that it requires more than 18 photons to be absorbed to excite the ground state by the IR pulse. Transitions of higher orders would thus be necessary to excite higher excited states directly from the IR pulse acting on the ground state. Therefore, it is a reasonable assumption to neglect the IR coupling of the ground state with excited states. Note that the weak effect of the IR field for intensities considered in this work is also suggested in \citab\cite{becker-faisal}. Therefore, the matrix elements of the dipole operator in the interaction matrix for the IR field $(\mathbf{d}_z^I)_{00L'n'}$ were explicitly set to zero (considering the ground state the first element in the basis set). This assumption allows for a dramatic improvement in the convergence of the solutions of the TDSE  with respect to the size of the basis set. On the other hand, it is the primary role of the XUV probe laser pulse to couple the ground state with the autoionizing state \sttstp and with nearby continuum states. Since it has rather high photon energy and low intensity (perturbative regime), it is unlikely that the XUV coupling between the doubly-excited states and the continuum states will play any significant role. Therefore, in the present study all the dipole matrix elements in the XUV interaction matrix $\mathbf{D}_X(t)$ except those coupling to the ground state $(\mathbf{d}_z^X)_{00L'n'}$, are explicitly set to zero. Note that a similar approximation for the XUV coupling in a similar context has been employed in \citab\cite{Wickenhauser}. Using this approximation the system of coupled differential equations~\eqref{eq:tdmat} can be written as follows:
  \begin{equation}
    i\mathbf{\dot{c}}(t)=\left[\mathbf{H}+\mathbf{V}_C+E_I(t+t_D)\mathbf{d}_z^I+E_X(t)\mathbf{d}_z^X\right]\mathbf{c}(t).
    \label{eq:approxmat}
  \end{equation}
  The structure of the matrices is showed in \figab\ref{fig:eqstr}.
  The atomic system is initially in the ground state. Considering the ground state to be the first element of the basis set, the initial condition can be written as $c_{0,1}(t_0)=1,\ c_{L,n>1}(t_0)=0$. The neglect of the ground state coupling with the other field-free eigenstates via the IR laser pulse allows for reduction of the total interval for the propagation of the TDSE as well. Since $\tau_I>\tau_X$, it would be necessary to start the propagation of the TDSE~\eqref{eq:tdmat} in the time $t_0=-t_D-\tau_I/2$, if  this assumption were not made. It would also then be necessary to perform the propagation of the wave function until the time $t_1=\max\left\{-t_D+\tau_I/2,\tau_X/2\right\}$. Although the total wave function described by the vector $\mathbf{c}(t)$ changes at times $t>t_1$ due to the presence of the CAP, the population of the ground state will remain constant after the end of the pulses, since the ground state wave function is well localized in the region where the CAP is zero. Therefore, for the purpose of the present project it is enough to perform the propagation until $t_1$ and extract the population of the ground state $c_{0,1}(t_1)$. The approximation that the IR field does not affect the ground state allows for the start of the propagation at time $t_0=-\tau_X/2$ and terminate it at time $t_1=\tau_X/2$. In view of the long duration of both pulses, this is an important simplification of the numerical calculations.
  
  The method of solution of the TDSE used in the present work is based on the split operator method in the basis set representation and it is described in the appendix in detail. This represents an alternative approach to the higher-order quadrature methods \cite{LagmagoKamta2002,nubbemeyer,Awasthi2005} frequently used in context of the interaction of atoms and molecules with strong laser fields. Application of the split-operator technique is straightforward in case of the wave-packet propagation (for details see \citab\cite{Bayfield1999} and references therein). However, the exponential of the time dependent operator remains a challenge in the basis set representation . \citet{Palacios2007} introduced an implementation in the DVR basis set which has properties that allow for an efficient numerical calculation of these exponentials. In the implementation used here these time-dependent exponentials are calculated in the diagonal representation of the dipole operators, as is discussed in the Appendix.
  
  
  The photoabsorption yields obtained from the solutions of the TDSE~\eqref{eq:approxmat} together with the static photoabsorption yield enable a calculation of the transient absorption spectra. This allows to relate the theoretical results presented here with the experimental transient absorption spectra published in \citab\cite{Loh2008}.

  \section{Field-free Hamiltonian and static photoabsorption spectrum}
  \label{sec:staticfree}
  Although we have found that the eigenrepresentation of the atomic Hamiltonian $H$ without the CAP is more practical for the solutions of the TDSE~\eqref{eq:approxmat}, it is useful to study the spectrum of the complex symmetric Hamiltonian $H_C=H+V_C$.
  
  This representation allows for easier optimization of the CAP parameter $\eta$ and of the size of the radial box $r_0$. The complex eigenenergies allow for a better insight into the structure of the discretized continuum and evaluate its quality for the solution of the full TDSE as well. The eigenfunctions $\phi_{Ln}(\mathbf{r}_1,\mathbf{r}_2)$ of the complex symmetric operator $H_C=H+V_C$ are given by the equation
  \begin{equation}
    H_C\phi_{Ln}(\mathbf{r}_1,\mathbf{r}_2)=E_{Ln}\phi_{Ln}(\mathbf{r}_1,\mathbf{r}_2),
  \end{equation}
  where the indices $L,n$ have identical meaning as in \eqab\eqref{eq:eigstates}. The complex eigenergies can be written as $E_{Ln}=E^r_{Ln}-i\Gamma_{Ln}/2$. They should reproduce well the energy positions $E_{Ln}^r$ and widths $\Gamma_{Ln}$ of the \sttstp and \sttpt autoionizing states for $L=0,1$. In addition, the density of the discretized continuum states must be sufficient to accurately describe the interaction between the autoionizing states and the continuum and to allow for converged calculation of the photoionization yield. These properties of the basis set are controlled by the size of the box $r_0$, the position of the absorbing boundary $r_b$ and the strength of the CAP $\eta$, as is described in \citab~\cite{Riss1993}. These parameters have been optimized in the present calculations to the values $r_0=200$\thinspace a.u. and $r_b=150$\thinspace a.u. These values allow for a good representation of the complex eigenenergies of the autoionizing states, they yield negligible decay widths of the ground and low excited states, and they provide a sufficient number of continuum states above the ionization threshold. The radial box is spanned by 700 \textsl{B}-splines of the sixth order. The \textsl{LS}-coupled expansion includes the single-particle partial waves up to $l=19$, every partial wave includes 200 radial orbitals (see \eqab\eqref{eq:oneorbs}). In order to describe the autoionizing states of the interest properly, the CI expansions of the states with $L=0$ and $L=1$ include the doubly-excited configurations, where both electrons can occupy orbitals with $l\leq 6$ and $j\leq6$. In the remaining configurations one electron is restricted to occupy a 1s, 2s, or 2p orbital, while the other electron can occupy any orbital, constrained so that the configurations contribute to the states with given spin $S$, total angular momentum $L$ and parity $P$. The higher partial waves $l$ included in the expansion are required for the representation of the continuum states having high total angular momentum $L$; they are essential for obtaining converged solutions of the full TDSE including the nonperturbative IR laser pulse.
    Optimization of the parameter $\eta$ as is described in the \citab\cite{Riss1993} using this basis set yields the value $\eta=10^{-4}$. The complex energies of the autoionizing states calculated using the basis set presented here are compared with previously published experimental results~\cite{Domke1996,Burgers1995} in \tabab\ref{tab:respos}.
  \begin{table}[htb]
        \caption{Comparison of the positions and widths of the \sttstp and \sttpt doubly-excited states calculated from diagonalization of the Hamiltonian $\mathbf{H}_C$ presented in this work with experimental values. The value $\eta=10^{-4}$ was used for the CAP parameter. The positions are relative to the ground state 1s$^2$. The values in brackets are obtained by fitting the perturbation calculations to the Fano line-shape.}
    \label{tab:respos}
    \begin{center}
      \begin{ruledtabular}
        \begin{tabular}{@{}dcdcd}
          & \multicolumn{2}{c}{Position (eV)} & \multicolumn{2}{c}{Width (eV)}\\
          \hline
          & \multicolumn{1}{l}{Present calculation} & \multicolumn{1}{l}{Experiment} & \multicolumn{1}{l}{Present calculation} & \multicolumn{1}{l}{Experiment}\\
          \multicolumn{1}{l}{\sttstp} & 59.687\ (59.688) & 60.15\text{ \cite{Domke1996}} & 0.0411\ (0.0456)& 0.0373\text{ \cite{Domke1996}}\\
          \multicolumn{1}{l}{\sttpt} & 61.747 & 62.06\text{ \cite{Burgers1995}} & 0.0063 & 0.0059\text{ \cite{Burgers1995}}\\
        \end{tabular}
      \end{ruledtabular}
    \end{center}
  \end{table}
  The widths of both autoionizing states calculated here are in good agreement with the experimental results. The discrepancies in their relative positions with respect to the ground state are mainly due to the inaccurate representation of the \stgnd ground state by our truncated atomic Hamiltonian $\mathbf{H}$. It gives a computed ground state energy of $E_{\mathrm{1s^2}}=-2.88343$\thinspace a.u., while the accurate experimental value is $E_{\mathrm{1s^2}}^{\mathrm{EXP}}=-2.90378$\thinspace a.u.~\cite{Bashkin1975}. To improve the ground state energy (and subsequently the relative positions of the double excited states) a significantly larger basis set would be necessary. However, the difference of doubly-excited state energies is more important for the calculations of the photoabsorption spectra than their relative position with respect to the ground state. As can be seen from \tabab\ref{tab:respos}, the energy difference according to the calculations is $\Delta E=2.06$\thinspace eV, while the experimental results give $\Delta E^\mathrm{EXP}=1.91$\thinspace eV.
  
\begin{figure}
    \includegraphics{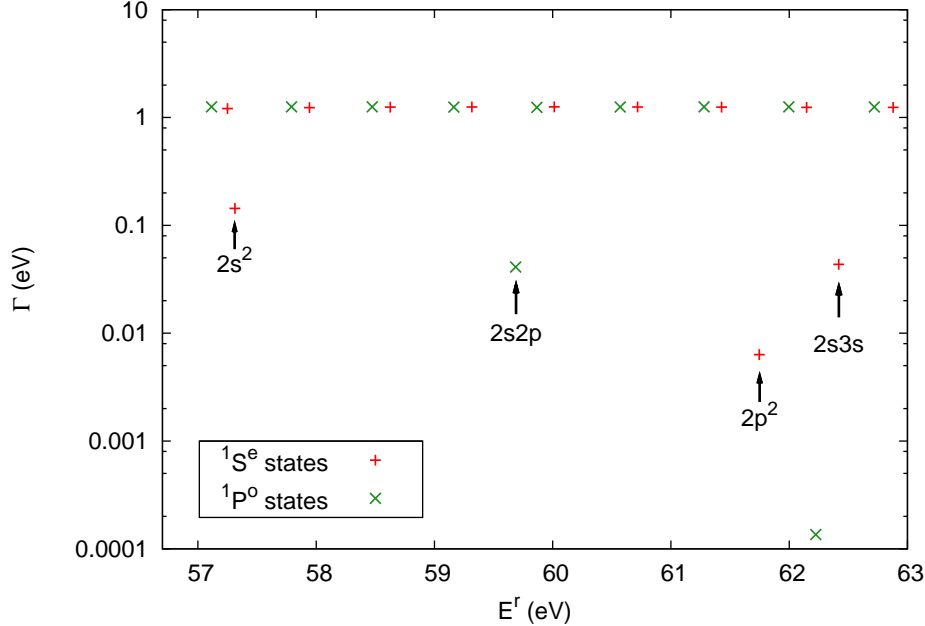}
        \caption{(Color online) Complex eigenenergies of the complex symmetric field-free Hamiltonian $\mathbf{H}_C$ spanning the interval of energies probed by the XUV laser pulse. Energies of states with the total angular momentum $L=0$ and $L=1$ are plotted. The autoionizing states are labeled by their principal independent particle CI configurations. The autoionizing state with energy 62.22\thinspace eV without label does not have a single dominant CI configuration, its discrete component is a linear combination of the 2s3p and 3s2p configurations. The $N=2$ ionization threshold is at energy 64.86\thinspace eV.}
    \label{fig:constat}
  \end{figure}

  The distribution of the $^1$S$^e$ and $^1$P$^o$ continuum states in the complex energy domain is plotted in \figab\ref{fig:constat}. The IR laser pulse assumed here allows for the optical field ionization of the doubly-excited states into the $N=2$ continuum. Since \citet{Loh2008} suggested that inclusion of this effect is important for the proper treatment of the phenomenon studied here, it is important to have a sufficiently high density of continuum states with higher total angular momenta $L$ in the energy range above the $N=2$ ionization threshold.
  Present calculations of the dipole matrix $\mathbf{d}_z^I$ yields $\mu_{ab}=2.11$\thinspace a.u. for the matrix element coupling the \sttstp and \sttpt states. This is in a good agreement with previously published value 2.17\thinspace a.u. obtained from the eigenchannel \textsl{R}-matrix calculations \cite{Loh2008}.
  
\begin{figure}
      \caption{(Color online) Loss of the ground state population in the absence of the IR pulse and its comparison with the Fano line-shape convolved with the bandwidth of the pulse.}
    \label{fig:statlos}
    \includegraphics{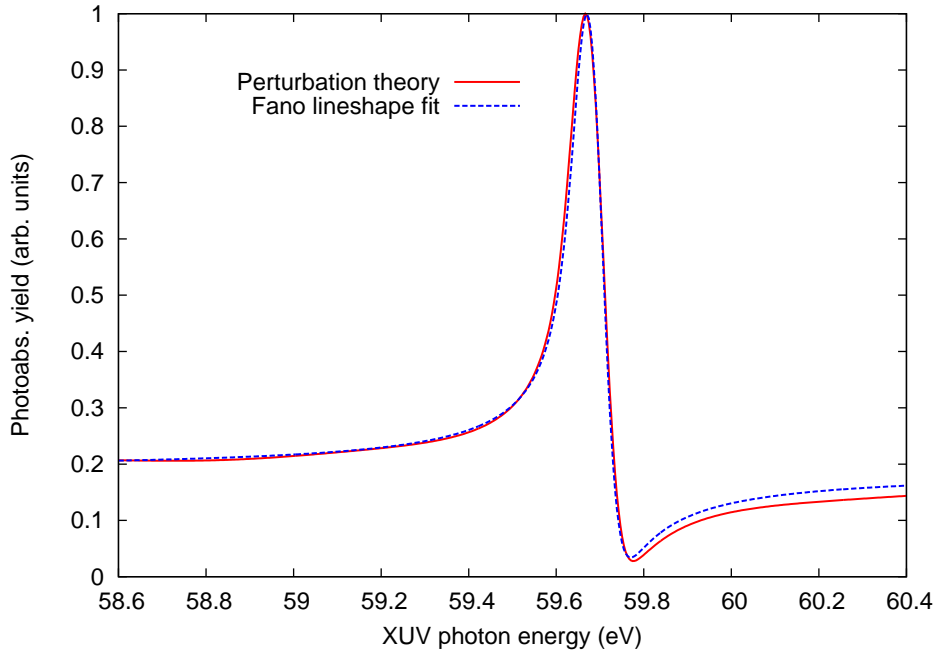}
  \end{figure}

  To understand the effect of the IR laser pulse it is a necessary prerequisite to calculate the static photoabsorption spectrum in the absence of the dressing laser pulse. The static photoabsorption yield corresponds to a Fano line-shape~\cite{fano-ci} across the known energy $\omega_1$ of the \sttstp doubly-excited state with respect to the \stgnd ground state, which has a known width $\Gamma_1$ and value $q_1$ of the Fano line-shape parameter. However, it is a stringent test of our basis set quality and of the CAP used throughout this work to calculate this photoabsorption yield independently using the present Hamiltonian and compare the corresponding spectrum with previously published results~\cite{Loh2008}. This can be done by setting $E_I^0=0$ in \eqab\eqref{eq:approxmat}. Taking into account the low intensity of the XUV laser field, it is most tractable to calculate the static photoabsorption spectrum using time-dependent perturbation theory. The corresponding loss of population of the ground state is shown in \figab\ref{fig:statlos}.
  This curve should correspond to the convolution of the Fano line-shape  with the the XUV laser pulse bandwidth. Therefore, we have fitted the absorption yield obtained from the perturbation calculations to this convolution and extracted the position, width and line-shape parameter of the \sttstp autoionizing state. The fitting procedure assumes a gaussian envelope of the laser pulse with bandwidth 0.06\thinspace eV (FWHM). The values for position and width obtained from the fitting are compared with previously published results in \tabab\ref{tab:respos}. As expected, they agree very well with the values obtained from diagonalization of $\mathbf{H}_C$. In addition, the fitting procedure yields $q_1=-2.71$, in good correspondence with the value -2.75 published in \citab\cite{Loh2008}. The convolved Fano line-shape corresponding to the fitted parameters is also shown in \figab\ref{fig:statlos}. The good correspondence between the perturbation calculations (using the $\cos^2$ XUV pulse envelope) and the Fano line-shape fit convolved with the gaussian function suggests that the results are not sensitive to the details of the XUV pulse envelope.
  
  \section{Photoabsorption by the dressed helium atom}
  \label{sec:dressed}
\begin{figure}
    \includegraphics{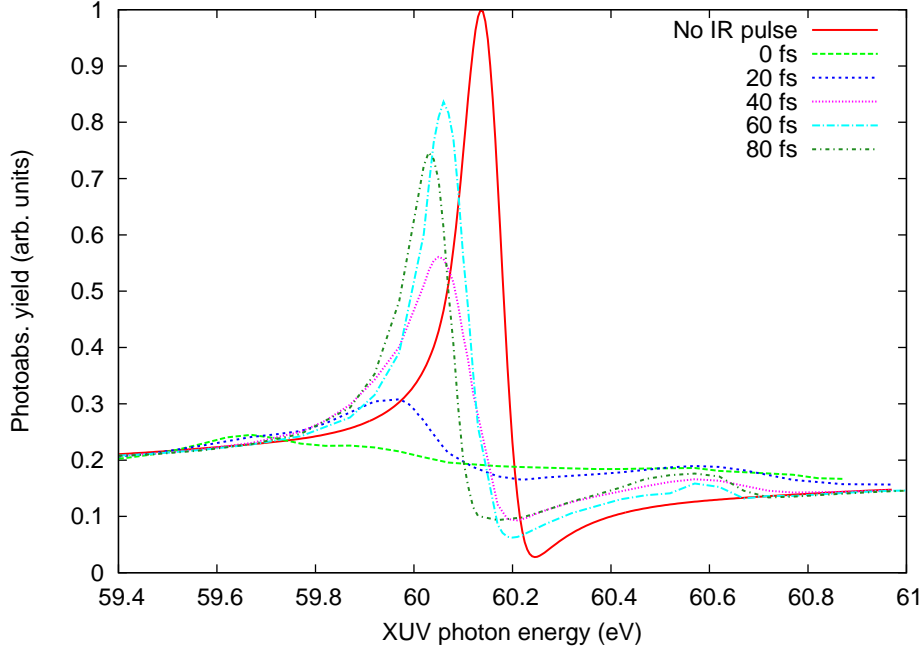}
        \caption{(Color online) Loss of the ground state population for several time delays as a function of the XUV photon energy calculated for the IR laser pulse with $\lambda_{\mathrm{IR}}=800$\thinspace nm and peak intensity $I_I^0=1.4\times 10^{13}$\thinspace W/cm$^2$}
    \label{fig:lospop}
  \end{figure}

  The loss of the ground state population at the end of the probe pulse $p(\omega_X)=1-\left|c_{00}(t_1)\right|^2$ calculated from \eqab\eqref{eq:approxmat} as a function of the XUV photon energy $\omega_X$ for several time delays $t_D$ is shown in \figab\ref{fig:lospop}.
  Since the duration of the dressing laser pulse is longer than the duration of the XUV pulse, different intensities of the IR field are probed at different time delays $t_D$. As can be seen in \figab\ref{fig:lospop}, the Fano line-shape disappears in the case of the perfect overlap of both pulses and the photoabsorption yield becomes essentially structureless. The peak and the minimum shifted with respect to the static case become more pronounced with increasing time delay $t_D$. An additional peak appears around the XUV photon energy 60.6\thinspace eV for $t_D>0$\thinspace fs. Its energy position does not change significantly with changing time delay. \citet{Loh2008} performed a model calculation of the photoabsorption yield based on the three-level model developed in \citab\cite{Madsden2000} describing the Autler-Townes doublet formation. It explains the presence of the additional peak in the photoabsorption yield as well as the shift of the higher peak with respect to the static case. The \citab\cite{Loh2008} model calculations that include optical field ionization of the doubly-excited states into the $N=2$ continuum due to the IR field confirm that this effect leads to a significant line-width broadening. This broadening makes the Autler-Townes doublet nearly invisible and the position of the peak at positive XUV detunings hardly shifts as the IR pulse intensity is changed. Since the calculations presented in this work were performed at an intensity that allows for optical field ionization, our results are in good correspondence with the theoretical considerations presented in \citab\cite{Loh2008}.
  
  The present calculations account for the time profile of both pulses, whereby the photoabsorption yields shown in \figab\ref{fig:lospop} together with the static photoabsorption yield plotted in \figab\ref{fig:statlos}, enable a calculation of the transient absorption spectra that can be compared with experimental results published in \cite{Loh2008}.
  
\begin{figure}
    \includegraphics{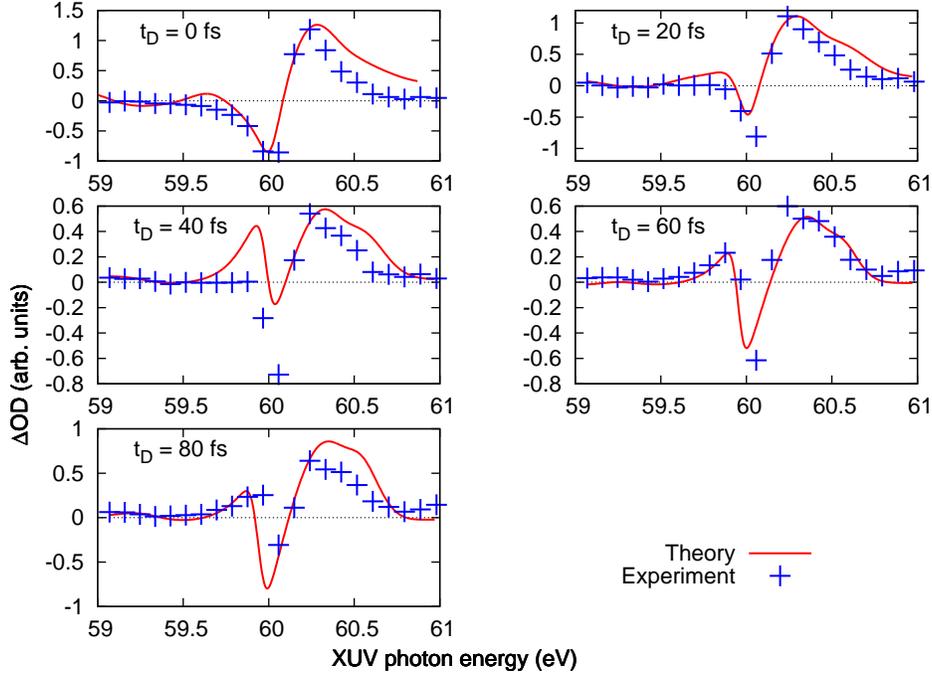}
        \caption{(Color online) Transient absorption spectra calculated using \eqab\eqref{eq:transab} for different time delays $t_D$ and their comparison with experimental results~\cite{Loh2008}. The value of the scaling constant $A=3.5$ was used to calculate the spectra with time delays 0\thinspace fs, 20\thinspace fs and 40\thinspace fs. The value $A=5$ was used to calculate the spectra with time delays 60\thinspace fs and 80\thinspace fs.}
    \label{fig:expercompar}
  \end{figure}

  The change of the optical density $\Delta OD$ is calculated using the formula that takes into account the energy resolution of the spectrometer used in the experiment~\cite{Loh2008}:
  \begin{equation}
    \Delta OD(\omega_X)=-\log_{10}\frac{\int\exp\left[-\sigma(\omega)n_t\ell-\frac{4\ln 2}{\Delta\omega^2}\left(\omega-\omega_X\right)^2\right]d\omega}{\int\exp\left[-\sigma_0(\omega)n_t\ell-\frac{4\ln 2}{\Delta\omega^2}\left(\omega-\omega_X\right)^2\right]d\omega},
    \label{eq:transab}
  \end{equation}
  where $\sigma(\omega)$ and $\sigma_0(\omega)$ are the photoabsorption cross section of the dressed medium and static photoabsorption cross section in the absence of the IR pulse, respectively. $n_t$ is the atom number density of the helium gas in the tube and $\ell$ is the length of the tube in the direction parallel to the direction of laser propagation. $\Delta\omega=0.18$\thinspace eV is the energy resolution of the spectrometer (FWHM) used in the experiment \cite{Loh2008}.
  Determination of the photoabsorption cross section from the photoionization yield obtained from \eqab\eqref{eq:approxmat} is not straightforward. The usual relation between the single-photon absorption cross section and the photoabsorption rate (probability per photon per second) $dp/dt$ for an infinitely long monochromatic pulse with photon energy $\omega$ is given by the formula
  \begin{equation}
    \sigma(\omega)=\frac{dp(\omega)}{dt}\frac{\omega}{I},
    \label{eq:csgeneral}
  \end{equation}
  where $dp/dt$ is the photoabsorption rate as a function of the photon energy $\omega$ and $I$ is the corresponding light intensity. This formula assumes that the photoabsorption rate does not depend on time. The situation in the present work is more complicated. Both pulses have a finite duration, and the photoabsorption rate varies with time. This time dependence should be included in \eqab\eqref{eq:csgeneral}. However, when one takes into account other assumptions made regarding the shape of the XUV pulse, it is a good approximation to calculate the photoabsorption cross section using the formula
  \begin{equation}
    \sigma(\omega_X)=A\frac{p(\omega_X)\omega_X}{T_XI_X},
    \label{eq:csformula}
  \end{equation}
  where $p(\omega_X)$ is the photoabsorption probability obtained from \eqab\eqref{eq:approxmat}, $T_X$ is the duration of the XUV pulse and $I_X$ is the corresponding XUV pulse intensity used to calculate $p(\omega_X)$. The scaling factor $A$ approximately accounts for the time dependence of the XUV field intensity and of the photoabsorption rate. Its value was chosen so that the calculated absorption spectra reproduce well the experimental transient absorption spectra published in \citab\cite{Loh2008}. Since $A$ is related to the XUV pulse only, the value used to calculate the static optical density is the same as the value used to calculate the optical density in the presence of the IR field. The comparison between our theoretical results and the \citab\cite{Loh2008} experiment is presented in \figab\ref{fig:expercompar}.
  The best correspondence between the present calclations and experimental results was achieved by setting $A=3.5$ for the time delays $t_D=0$\thinspace fs, 20\thinspace fs and 40\thinspace fs. The remaining spectra with time delays $t_D=60$\thinspace fs and 80\thinspace fs compare best with the experimental results for $A=5$. Therefore, the best fit of the experimental data is obtained when the product $I_XT_X$ used in the calculations is smaller than the values estimated in the experimental work \cite{Loh2008}. The calculated transient absorption spectra plotted in \figab\ref{fig:expercompar} have been shifted on the energy axis by 0.45\thinspace eV to correct for the inaccuracy of the \sttstp doubly-excited state position with respect to the ground state in the present calculations (see \tabab\ref{tab:respos} and the discussion in \secab\ref{sec:staticfree}). A change of the value of the parameter $A$ affects the ratio of the peak magnitudes in the transient absorption spectra, however it does not affect the positions of the structures in the curves. As can be seen in \figab\ref{fig:expercompar}, the present theoretical calculations are generally in encouraging agreement with the experimental results. The left peak that develops with  increasing time delay is a consequence of the rising maximum of the photoabsorption spectrum due to decreasing IR intensities probed by the XUV pulse. The figure shows that this peak develops in the present calculations more rapidly with increasing time delay than is evident in the experimental results. The peak at positive XUV detunings calculated in the present study corresponds well to the experimental results. Since the probe pulse is produced by the HHG, it is expectable that its spatial and frequency profile have more difficult structure than a simple gaussian envelope as it is assumed in this work. Therefore, the discrepancies between the experimental transient absorption spectra published in \citab\cite{Loh2008} and the spectra presented here can possibly be attributed to the simplified shape of the XUV pulse assumed in the present theoretical treatment.
  
\begin{figure}
    \includegraphics{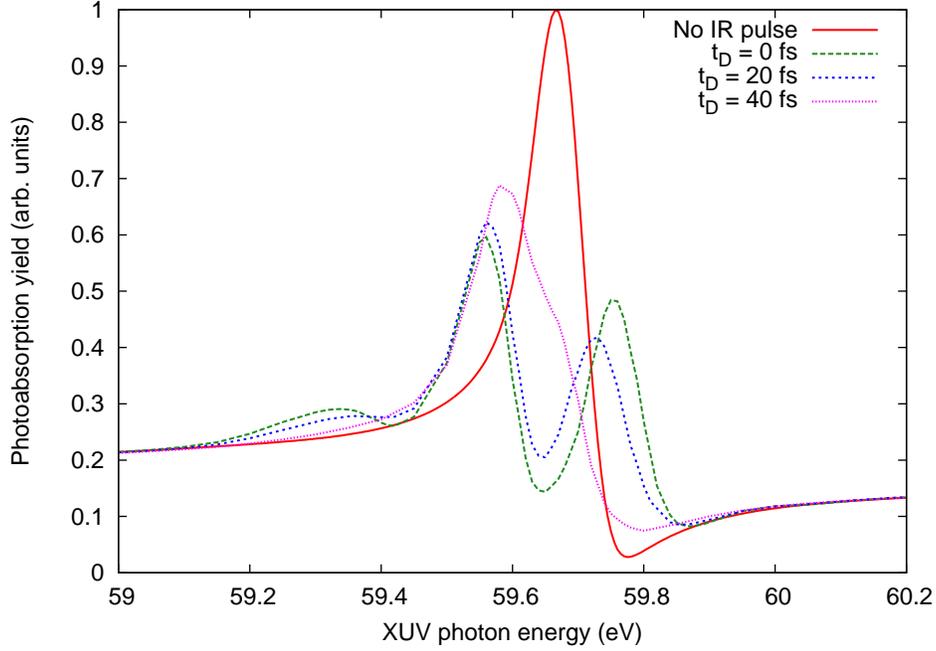}
        \caption{(Color online) Photoabsorption yield calculated for the peak intensity of the IR field $I_0=7\times 10^{11}$\thinspace W/cm$^2$ and for the dressing pulse wavelength $\lambda_I=600$\thinspace nm corresponding to the energy difference between the \sttstp and \sttpt doubly-excited states. Curves for time delays $t_D=0$\thinspace fs, 20\thinspace fs 40\thinspace fs are compared with the static photoabsorption yield with no IR pulse}
    \label{fig:resonir-weak}
  \end{figure}

  Although the photoabsorption yields in \figab\ref{fig:lospop} show the reduced XUV photon absorption caused by the dressing IR laser pulse, the relation between these results and the EIT-like structures is not straightforward. The spectral peaks corresponding to the dressed states of the Autler-Townes doublet approach each other linearly as the Rabi splitting $\Omega_c$ of the states coupled by the IR laser pulse decreases. This behavior is not apparent in \figab\ref{fig:lospop}. The Rabi splittings $\Omega_c$ probed by the peak of the XUV laser pulse at time delays $t_D=0$\thinspace fs, 20\thinspace fs and 40\thinspace fs are $\Omega_c=1.18$\thinspace eV, 0.62\thinspace eV and 0.0025\thinspace eV, respectively. The splitting for larger time delays $t_D$ is significantly smaller than the width of the \sttstp doubly-excited state. However, \figab\ref{fig:lospop} does not show any corresponding rapid change in the energy separation of the peaks as the time delay is changed. The clear interpretation of these photoabsorption yields is complicated by the optical field ionization of the doubly-excited states, by the finite durations of the laser pulses, and by the large detuning of the dressing laser from the energy difference of the \sttstp and \sttpt doubly-excited states. In order to eliminate the effect of the large IR detuning, another calculation was performed with the wavelength of the IR laser pulse modified to 600\thinspace nm. The corresponding photon energy 2.06\thinspace eV now matches the calculated energy difference between the doubly-excited states \sttstp and \sttpt (see \tabab\ref{tab:respos}). In order to suppress the effect of the optical field ionization, the peak intensity of the coupling IR laser pulse was decreased to $I_I^0=7\times 10^{11}$\thinspace W/cm$^2$. Corresponding photoabsorption yields are shown in \figab\ref{fig:resonir-weak}.
  This graph shows for $t_D=0$\thinspace fs two peaks with energy separation 0.2\thinspace eV and minimum between them at energy close to the position of the static photoabsorption peak. The Rabi splitting of the autoionizing states coupled by the IR laser pulse with the reduced peak intensity used here is $\Omega_c=0.27$\thinspace eV. The curve for time delay $t_D=20$\thinspace fs shows similar behavior. The energy separation of the peaks is 0.16\thinspace eV, the Rabi splitting of the \sttstp and \sttpt doubly-excited states probed by the peak of the XUV laser pulse is $\Omega_c=0.145$\thinspace eV. In case of larger time delay $t_D=40$\thinspace fs both peaks have a strong overlap and their energy separation is smaller than the width of the \sttstp doubly-excited state. This correspondence confirms that we can interpret these structures as the Autler-Townes doublet. Energy separations of the peaks for single time delays $t_D$ do not exactly match the corresponding Rabi splittings $\Omega_c$, mainly because of the finite duration of both pulses. Another reason could be the coupling of additional doubly-excited states by the IR laser pulse that would add further complications to the three-level picture. The presence of an additional autoionizing state in the coupling scheme also appears to explain the smallest additional peak at an energy below 59.4\thinspace eV which occurs at small time delays in \figab\ref{fig:resonir-weak}. The \sttst doubly-excited state is a good candidate, since the energy difference between the \sttst and \sttstp doubly-excited states is 2.373\thinspace eV in the present calculations.  In general, the calculations performed with the reduced IR intensity and with the photon energy adjusted as discussed above show the EIT-like structure more clearly than the calculations performed with parameters of the IR laser pulse used in the experiment. This suggests that the Autler-Townes structure in the calculations carried out using the experimental parameters are not very pronounced simply because the dressing laser detuning is so large.
  
\begin{figure}
    \includegraphics{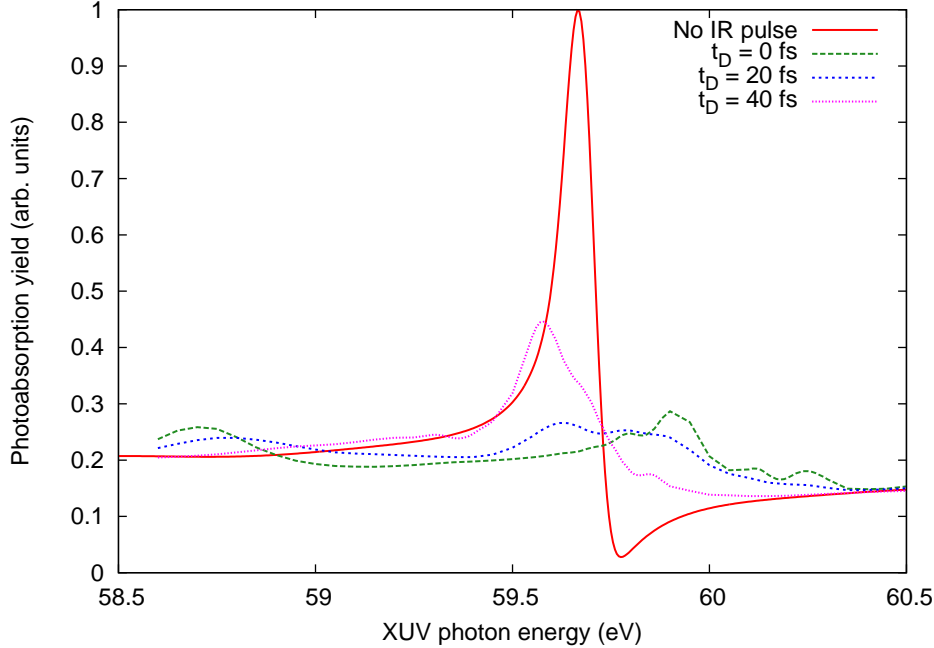}
        \caption{(Color online) The photoabsorption yield calculated for the peak IR field intensity $I_0=1.4\times 10^{13}$\thinspace W/cm$^2$ and for the dressing pulse wavelength $\lambda_I=600$\thinspace nm corresponding to the energy difference between the \sttstp and \sttpt doubly-excited states. Curves for time delays $t_D=0$\thinspace fs, 20\thinspace fs 40\thinspace fs are compared with the static photoabsorption yield.}
    \label{fig:resonir-strong}
  \end{figure}

  In order to investigate the manner in which the optical field ionization of the autoionizing states affects the dressed photoabsorption spectra, another calculation with the adjusted IR laser pulse ($\lambda_I=600$\thinspace nm) was performed. The experimental peak intensity of the IR dressing laser pulse $I_I^0=1.4\times 10^{13}$\thinspace W/cm$^2$ was used. The resulting photoabsorption yields for several time delays $t_D$ are shown in \figab\ref{fig:resonir-strong}.
  This graph shows a broad peak around the energy 58.7\thinspace eV and a richer structure in the energy interval from 59.7\thinspace eV to 60.3\thinspace eV in case of perfect temporal overlap of the two laser pulses. These structures approach each other with increasing time delay $t_D$. Since optical field ionization broadens the line-widths, the characterization of the individual peaks in the photoabsorption spectra is not straightforward. However, the well pronounced separation of the broad peak at negative XUV detuning and the structure at positive XUV detuning is most likely related to the coupling of the \sttstp and \sttpt autoionizing states, as can be assumed from the change of the separation with increasing time delay $t_D$. It is possible that the IR field at the experimental peak intensity $I_I^0=1.4\times 10^{13}$\thinspace W/cm$^2$ enables coupling of additional autoionizing states and this coupling leads to the rich structure of peaks appearing at positive XUV detunings.
  \section{Conclusions}
  This work presents theoretical treatment of the photoabsorption in XUV for helium dressed by the IR laser. The \sttstp and \sttpt doubly-excited states are coupled by strong 800\thinspace nm laser pulse with duration 42\thinspace fs. This system is probed by the weak XUV pulse produced by HHG. The photoabsorption yields in the presence of both pulses are calculated as a function of the XUV photon energy for several dressing-probe time delays by solving the TDSE. The calculated transient absorption spectra are in reasonable correspondence with experimental results of \citet{Loh2008}. The presence of the coupling IR laser significantly reduces the peak photoabsorption yield and changes of the time delay between the dressing and probe pulse show the dependence on the IR field intensity that is probed by the XUV photons. Calculations performed with the wavelength of the dressing laser modified to $\lambda_I=600$\thinspace nm, corresponding to zero detuning of the IR laser, clearly show a structure in the photoabsorption spectrum that can be interpreted as the Autler-Townes doublet and an EIT-like structure. This allows us to conclude that this was not clearly pronounced in the experimental results published in \citab\citet{Loh2008} primarily because of the large detuning of the IR laser. The photoabsorption yields calculated for different dressing laser intensities showed that the optical field ionization of the doubly-excited states broadens the linewidths and increasing dressing laser intensity leads to formation of more complicated structures in the photoabsorption spectra. These could result from the increasing importance of additional autoionizing states at higher IR intensity.
  
  The method of solution of the TDSE presented in this work proved to be stable and suitable for theoretical treatment of two active electron atoms in pump-probe settings of the laser pulses with a potential of further improvement in the future.
  \begin{acknowledgments}
    Conversations with Zhi-Heng Loh and Stephen Leone are greatly appreciated, as is access to some of their unpublished data and calculations. We also thank Andreas Becker and Antonio Pic\'on for stimulating discussions. This work was supported in part by the Department of Energy, Office of Science. This research used resources of the National Energy Research Scientific Computing Center, which is supported by the Office of Science of the U.S. Department of Energy.
  \end{acknowledgments}
  \appendix*
  \section{Method of the solution of the TDSE}
  \label{sec:app}
  Since it is necessary to integrate the system of equations~\eqref{eq:tdmat} for a set of different XUV photon energies $\hbar\omega_X$ and pump-probe time delays $t_D$, it is important to employ an integration method that efficiently meets these requirements. After tests of several different methods used elsewhere in similar contexts (e.g. Runge-Kutta~\cite{LagmagoKamta2002} and Crank-Nicholson~\cite{Bayfield1999}) we chose to implement the split-operator technique~\cite{Bayfield1999} in a basis set representation. The basic idea of this propagation method is that if the total time-dependent Hamiltonian $\mathbf{H}_T(t)$ can be written as a sum of the time-independent and time-dependent parts $\mathbf{H}_T(t)=\mathbf{H}_0+\mathbf{H}_1(t)$, the propagation of the wave function between times $t$ and $t+\Delta t$, where $\Delta t$ is sufficiently small, can be implemented to second order in the time step, as follows:
  \begin{equation}
    \mathbf{\Psi}(t+\Delta t)=\exp\left(-\frac{i\Delta t}{2}\mathbf{H}_0\right)\exp\left[-i\int_t^{t+\Delta t}\mathbf{H}_1(t')dt'\right]\exp\left(-\frac{i\Delta t}{2}\mathbf{H}_0\right)\mathbf{\Psi}(t)+\mathbf{O}\left[\left(\Delta t\right)^3\right].
    \label{eq:splitop}
  \end{equation}
  If we denote
  \begin{equation}
    F_{I,X}(t,\Delta t)=\int_t^{t+\Delta t}E_{I,X}(t')dt',
  \end{equation}
  then the application of this method to the present system of equations~\eqref{eq:approxmat} yields
  \begin{eqnarray}
    \mathbf{c}(t+\Delta t)=\exp\left[-\frac{i\Delta t}{2}\left(\mathbf{H}+\mathbf{V}_C\right)\right]
    \exp\left[-i\mathbf{d}_z^IF_I(t+t_D,\Delta t)-i\mathbf{d}_z^XF_X(t,\Delta t)\right]\nonumber\\
    \exp\left[-\frac{i\Delta t}{2}\left(\mathbf{H}+\mathbf{V}_C\right)\right]\mathbf{c}(t)+\mathbf{O}\left[\left(\Delta t\right)^3\right].
    \label{eq:basprop}
  \end{eqnarray}
  The first and third exponential terms in this equation can be calculated rather easily, as they are time independent (for a fixed time step $\Delta t$) and all the matrices are block diagonal. Since the eigenrepresentation of the field-free Hamiltonian $H$ is used, this term can be expressed using the Crank-Nicholson formula
  \begin{equation}
    \exp\left[-\frac{i\Delta t}{2}\left(\mathbf{H}+\mathbf{V}_C\right)\right]=\frac{\mathbf{1}-i\left(\mathbf{H}+\mathbf{V}_C\right)\Delta t/4}{\mathbf{1}+i\left(\mathbf{H}+\mathbf{V}_C\right)\Delta t/4}+\mathbf{O}\left[\left(\Delta t\right)^3\right],
    \label{eq:cnpart}
  \end{equation}
  which is accurate to second order in the time step. Although the evaluation of this term requires a matrix inversion, it is not demanding, as it can be performed independently for every block matrix corresponding to a definite angular momentum $L$. In addition, it is necessary to perform this step only once before calculating the time propagation, as this term is time independent.

  Calculation of the second exponential in \eqab\eqref{eq:basprop} is more difficult, as it explicitly depends on time and is not block diagonal. Since $\mathbf{d}_z^I$ and $\mathbf{d}_z^X$ do not commute (see the structure in \figab\ref{fig:eqstr}), we can write this term using the split-operator formula~\eqref{eq:splitop} as follows:
  \begin{eqnarray}
    \exp\left[-i\mathbf{d}_z^IF_I(t+t_D,\Delta t)-i\mathbf{d}_z^XF_X(t,\Delta t)\right]=\exp\left[-\frac{i}{2}\mathbf{d}_z^IF_I(t+t_D,\Delta t)\right]\nonumber\\
    \exp\left[-i\mathbf{d}_z^XF_X(t,\Delta t)\right]\exp\left[-\frac{i}{2}\mathbf{d}_z^IF_I(t+t_D,\Delta t)\right]+\mathbf{O}\left[\left(\Delta t\right)^3\right].
    \label{eq:secondsplit}
  \end{eqnarray}
  The most straightforward way to calculate these exponentials is using the eigenrepresentations of $\mathbf{d}_z^I$ and $\mathbf{d}_z^X$:
  \begin{eqnarray}
    \exp\left[-i\mathbf{d}_z^IF_I(t+t_D,\Delta t)-i\mathbf{d}_z^XF_X(t,\Delta t)\right]=\mathbf{U}_I\mathrm{diag}\left\{\exp\left[-\frac{i\lambda^I_k}{2}F_I(t+t_D,\Delta t)\right]\right\}\mathbf{U}_I^T\nonumber\\
    \mathbf{U}_X\mathrm{diag}\left\{\exp\left[-i\lambda^X_kF_X(t,\Delta t)\right]\right\}\mathbf{U}_X^T\mathbf{U}_I\mathrm{diag}\left\{\exp\left[-\frac{i\lambda^I_k}{2}F_I(t+t_D,\Delta t)\right]\right\}\mathbf{U}_I^T,
  \end{eqnarray}
  where $\lambda^{I,X}_k$ are the eigenvalues of $\mathbf{d}_z^{I,X}$ and $\mathbf{U}_{I,X}$ are the corresponding matrices of the column eigenvectors. Substitution of this result together with \eqab\eqref{eq:cnpart} into \eqab\eqref{eq:basprop} and denoting
  \begin{equation}
    \mathbf{V}_1=\mathbf{U}_I^T\frac{\mathbf{1}-i\left(\mathbf{H}+\mathbf{V}_C\right)\Delta t/4}{\mathbf{1}+i\left(\mathbf{H}+\mathbf{V}_C\right)\Delta t/4}\mathbf{U}_I,\qquad\mathbf{V}_2=\mathbf{U}_I^T\mathbf{U}_X,\qquad\mathbf{c}_I(t)=\mathbf{U}_I^T\mathbf{c}(t)
    \label{eq:propmatintro}
  \end{equation}
  yields
  \begin{eqnarray}
    \mathbf{c}_I(t+\Delta t)=\mathbf{V}_1\mathrm{diag}\left\{\exp\left[-\frac{i\lambda^I_k}{2}F_I(t+t_D,\Delta t)\right]\right\}\mathbf{V}_2\mathrm{diag}\left\{\exp\left[-i\lambda^X_kF_X(t,\Delta t)\right]\right\}\mathbf{V}_2^T\nonumber\\
    \mathrm{diag}\left\{\exp\left[-\frac{i\lambda^I_k}{2}F_I(t+t_D,\Delta t)\right]\right\}\mathbf{V}_1^T\mathbf{c}_I(t).\qquad
    \label{eq:irprop}
  \end{eqnarray}
  This propagation scheme in one time step means that the wave function is first expressed in the eigenrepresentation of $\mathbf{d}_z^I$ and propagation for a quarter of the time step is performed using the field-free Hamiltonian augmented by the CAP. Then the wave function is propagated in the IR field for half of the time step. After that the wave function is transformed into the eigenrepresentation of the XUV dipole operator and the propagation for one time step is performed in the XUV field only. The wave function is transformed back to the IR eigenrepresentation again and the propagation steps in the IR field only as well as the step using the field-free Hamiltonian are performed. This propagation scheme involves only two matrices which are time dependent and they are diagonal in different representations. It means that only two changes of representation are necessary in every time step. Note that application of \eqab\eqref{eq:irprop} in two subsequent time steps allows for multiplication by the matrix $\mathbf{V}_1^T\mathbf{V}_1$ between the steps. Therefore, every propagation step involves three matrix-vector multiplications and two diagonal matrix-vector multiplications, where only the diagonal matrices are time dependent. The matrix $\mathbf{V}_2$ is real and the matrix $\mathbf{V}^T_1\mathbf{V}_1$ is complex symmetric. After the transformation of the initial state vector $\mathbf{c}(t_0)$ from the energy eigenrepresentation of $\mathbf{H}$ to the diagonal representation of $\mathbf{d}_z^I$, the rest of the propagation is performed in this basis. This approach is permissible in the context of the problem studied here, because we only need to determine the depletion of the field-free ground state at the end of the XUV laser pulse only, rather than information about the evolution of the complete wave function during the pulse. Note that only the function $F_X(t,\Delta t)$ depends on the XUV photon energy $\hbar\omega_X$ and the information about the time delay $t_D$ appears only in the diagonal matrix in \eqab\eqref{eq:irprop}.
  Therefore, the same matrices $\mathbf{V}_1^T\mathbf{V}_1$  and $\mathbf{V}_2$ can be used in all the time dependent calculations for all the values of $\hbar\omega$ and $t_D$ of interest.
  
  One more note related to \eqab\eqref{eq:secondsplit} should be made at this point. The split operator method is used there to calculate the exponential involving matrices $\mathbf{d}_z^I$ and $\mathbf{d}_z^X$ which do not commute. The non-zero commutator is a consequence of the approximation introduced in \secab\ref{sec:theo}, where we neglect the effect of the IR laser field on the ground state and neglect the XUV couplings that do not involve the ground state (see \figab\ref{fig:eqstr}). Validity of this approximation implies that the commutator $\left[\mathbf{d}_z^I,\mathbf{d}_z^X\right]$ is sufficiently small to write the exponential on the left hand side of \eqab\eqref{eq:secondsplit} simply as a product of two exponentials. This would change the form of the final time propagator \eqref{eq:irprop} in such way that it will contain only two time-dependent matrices. Therefore, this approach would save one matrix-vector multiplication in every time step. However, its impact on the accuracy of the time propagation needs to be tested and its implementation will be the subject of forthcoming research.
  
  In conclusion, the time required to calculate the matrices $\mathbf{V}_1^T\mathbf{V}_1$ and $\mathbf{V}_2$ (matrix diagonalization, inversion and the matrix-matrix multiplication) is negligible compared to the calculations of the time propagation using \eqab\eqref{eq:irprop}. This propagation scheme is of the second order in the time step and scales as $N^2$ with the dimension of the basis set used (since it involves only matrix-vector multiplications in every time step).
%

\end{document}